# A Wrapper of PCI Express with FIFO Interfaces based on FPGA

Hu Li, Yuan'an Liu, Dongming Yuan, Hefei Hu
Wireless and EMC Laboratory
Beijing University of Posts and Telecommunications (BUPT)
Beijing, China
E-mail: sdbzlh@163.com, yuliu@bupt.edu.cn, yuandm@bupt.edu.cn, huhefei@bupt.edu.cn

*Abstract*—This paper proposes a PCI Express (PCIE) Wrapper core named PWrapper with FIFO interfaces. Compared with other PCIE solutions, PWrapper has several advantages such as flexibility, isolation of clock domain, etc. PWrapper is implemented and verified on Vertex-5-FX70T which is a development board provided by Xilinx Inc. Architecture of PWrapper and design of two key modules are illustrated, which timing optimization methods have been adopted. Then we explained the advantages and challenges of on-chip interfaces technology based on FIFOs. The verification results show that PWrapper can achieve the speed of 1.8Gbps (Giga bits per second).

## I. INTRODUCTION

As the development of electronic industry, data processing ability of a single chip increases. To satisfy the growing appetite for bandwidth, a new bus technology called PCI Express was introduced to replace PCI, PCI-X, and AGP. PCI Express offers a serial architecture that alleviates some of the limitations of parallel bus architectures by using clock data recovery (CDR) and differential signaling.[1]

PCI Express technology use packet-based transaction non-shared data bus, which achieves a very high speed of 2.5Gbps/lane/direction theoretically.[2] There are basically three layers introduced in PCI Express Specification, as shown in Fig. 1. The Transaction Layer is the upper one, whose primary function is to transmit and receive transaction layer packets (TLPs). The Data Link Layer is the middle one, whose primary responsibility is to provide a reliable mechanism for the exchange of TLPs. The Physical Layer interfaces the Data Link Layer with signaling technology for link data interchange and is responsible for framing, de-framing, scrambling, descrambling, 8b/10b encoding and decoding of TLPs and DLLPs.[2]

Nowadays, many companies provide PCI Express solutions, which can be classified into FPGA-based and ASIC-based according to implementation. The validity and compliance of these two kinds of solutions are unquestioned. But FPGA gives designers the ability to create a design that exactly matches their requirements and more flexibility. When using FPGA as the solution, there are two alternatives: one-chip solution and two-chip solution. The one-chip solution utilizes a high-performance FPGA to implement the entire protocol, while in the two-chip solution a low-cost FPGA is connected to a standalone PCI Express PHY over a PIPE interface.[1]

Among these solutions based on FPGA, [1] provided a design with SDRAMs interface, which requires SDRAM resources out of FPGA, [3] provided a design without explicit and simple interfaces for others to use. They either used extra resources not on chip or did not provide interfaces for others to reuse. To overcome these problems and to facilitate others to reuse our design, we propose architecture with FIFO interfaces in this paper, which aiming at achieving the raw data speed above 1.5Gbps.

In this paper, a Xilinx Virtex-5 FX70T chip is used to implement the one chip solution. And Xilinx Development Board ML507 is chosen as the design platform. There is at least one PCI Express hard core in each Virtex-5 chip except for a few types. And a wrapper core called Endpoint Block Plus for PCI Express (shorted as EBlock hereafter) is provided by Xilinx Inc. This two cores work together to implement the whole Physical Layer and Data Link Layer defined in PCI Express specification.[5]

An IP core named PCIE Wrapper (shorted as PWrapper hereafter) is proposed in this paper using these two cores, offering a flexible but simple interface using FIFOs. This architecture is illustrated in Section II, including advantages and challenges of on-chip interfaces technology based on FIFOs. Timing optimization is also provided in Section II. Resource usage and verification results are given in Section III. Conclusion is drawn finally in Section IV.

## II. DESIGN AND TIMING OPTIMIZATION

### A. Architecture

Overall architecture is composed of three parts, PWrapper, EBlock and the hard core of PCI-E, as shown in Fig. 2. PWrapper is made up of a transmission (TX) module, a reception (RX) module and a configuration (CONF) module. TX and RX modules are where most packet switching capabilities are integrated, and they connect to the RX and TX interfaces of EBlock. CONF&INTR Module is designed to configure EBlock and offers interrupt interface to other modules.

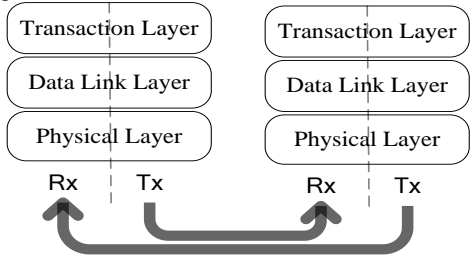

Figure 1. Three Layers of PCI Express Protocol.

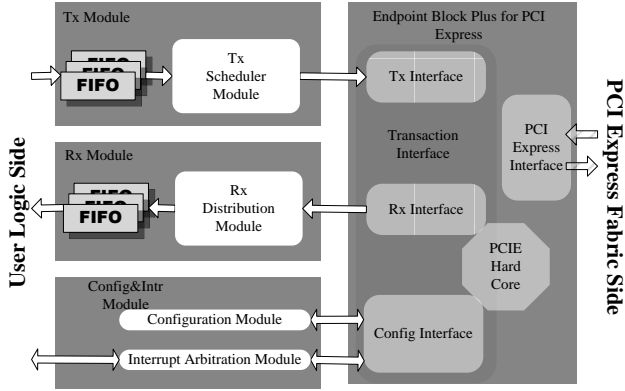

Figure 2. Architecture of PWrapper and EBlock.

### B. On-chip Interface Technology Using FIFOs

Communication method of different modules in one FPGA is of great importance, since its efficiency and complexity significantly affect the performance of the design. The simplest method to communicate is to use signals (wires inside FPGA), but when the design is separated into several clock domains this method fails. Other popular method is to use bus such as Wishbone, AMBA, or the modern one AXI, but the resources occupied by the bus may not be ignored if the FPGA design is specific and relatively small.

To offer a simple but flexible interface of PWrapper, a method using FIFOs is adopted, which means the communication between modules and PWrapper is accomplished by FIFOs. The function is similar to but simpler than the pipeline mode of Wishbone.

*1) Advantages:* The advantages of this method are shown as follows.

*a)* FIFO cores are simple but reliable, since almost all FPGA vendors provide specific and robust FIFO cores.

*b)* FIFOs can be used between 2 clock domains.

*c)* FIFOs can be used to convert data width.

*d)* FIFOs use the Brams inside FPGA. The amount of Brams is large in one chip, and therefore flip-flop resources can be saved.

*e)* FIFOs are flexible, since their parameters such as width and depth can be configured according to circumstances of the design, such as clock frequency, data rate, data width of the 2 modules, the Bram resources remaining in the chip and so on.

*2) Challenges:* However, there are also several challenges using this method.

*a)* The first challenge is that datum will be lost if be written into a full FIFO or read from an empty FIFO.

There are basically 3 approaches to deal with this challenge.

Approach #1 is to monitor the success indicating signals when writing and reading. Such signals are WR_ACK and Valid in Xilinx IP core. [5]

Approach #2 is to monitor the amount of datum in a FIFO, then write only when the space is adequate for one packet and read only when the datum is not less than one packet. To do this, both sides should get the information about the length of the packet which is usually included in the head of the packet. The two signals WR_DATA_COUNT and RD_DATA_COUNT, as well as the First-Word Fall-Through technique may be used here.

Approach #3 is to set read and write threshold for read and write operation, then read only when the data amount is not less than read threshold, write only when the data amount is not more than write threshold.

Approach #1 is simple, but doing this necessitates a much more complex FSM (Finite State Machine), which judge every time after read and write operation and go back to the previous state every time when an operation fails. Approach #2 reduces the complexity of FSM, which need to wait at the beginning of one operation but can accomplish one operation without interruption and back-off. Approach #3 simplified the waiting condition in approach #2 and is extremely suitable for a system transmitting packet the length of which is pre-determined.

*b)* The second challenge is based on the fact that, although FIFO core is reliable in most cases, if one packet is corrupted in the FIFO especially the length field of the packet is unfortunately destroyed, then all the following packets in the FIFO or going into the FIFO later will not be recognized. This disastrous situation will be not rare when the data speed is very high.

To deal with this challenge, extra bit is used to mark the start and end of one packet to reduce the data corruption into one or several packet. And CRC (cyclic redundancy check) is needed when the data comes through several FIFOs or from outside the chip.

*c)* The Third challenge is to decide the values of parameters of FIFOs.

Here trade-off is needed. Because the wider and the deeper the FIFOs are, the more resources are used. But a larger width will reduce the clock frequency constraint of the modules attached to the FIFO when the data rate is constant. And a larger depth can facilitate system to process data stream with a higher speed. So these two parameters should be considered according to the situation.

### C. Design of RX Module & TX Module

*1) RX Module:* RX module is responsible for receiving and distributing incoming packets. It provides a simple but effect interface (standard FIFO interface) to modules that need to receive TLPs. Incoming TLPs are detected by the RX module by inspecting the signals given out by EBlock , and then travel through RX module to its destination.

RX module consists of a RX Handler module and several (at least one) First-In First-Out (FIFO) queues as shown in Fig. 2.

RX Handler is implemented using a single Finite State Machine (FSM). Simplified State Diagram of RX FSM is shown in Fig. 3. Once a TLP is detected on the interface of EBlock, RX Handler reads the first two QW of this TLP to get the destination address (hereafter shorted as DA) of it. Then RX FSM stores this TLP to corresponding FIFO according to DA. If no FIFO is responsible for this TLP, it will be discarded or stored in a special FIFO for further processing. RX FSM meets all the timing relationship of EBlock defined in [4], so it can support Basic TLP Receive Operation, Data Path Throttling and Back-To-Back Transactions for higher performance. RX FSM also meets all the timing relationship of FIFO defined in [5].

*2) TX Module:* TX module is responsible for scheduling and transmitting packets. Same as RX module, it provides FIFO interface to modules that need to transmit TLPs. All outgoing TLPs are stored in FIFOs, and then sent out one by one to EBlock according to predetermined priority.

TX module consists of a TX Scheduler module and several (at least one) FIFO queues as shown in Fig. 2.

TX Scheduler module is implemented using two FSMs, Judging FSM and TX FSM. Judging FSM is responsible for judging which TLP will be sent at present according to the priority of FIFOs and other factors. TX FSM is responsible for transmitting TLPs from FIFO to EBlock with the assistance of Judging FSM. Simplified State Diagrams of these two are shown in Fig. 4. Once there are integrate TLPs detected by Judging FSM using read counter of FIFO, it decides the transmitting order and informs TX FSM about this order. TX FSM then sends the packet out to EBlock and informs Judging FSM.

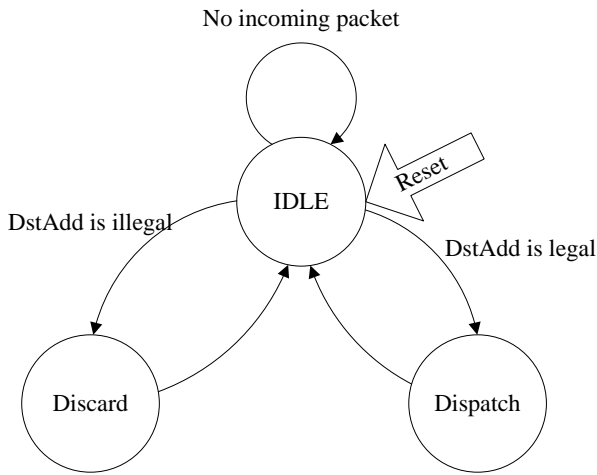

Figure 3. Simplified State Diagram of RX FSM.

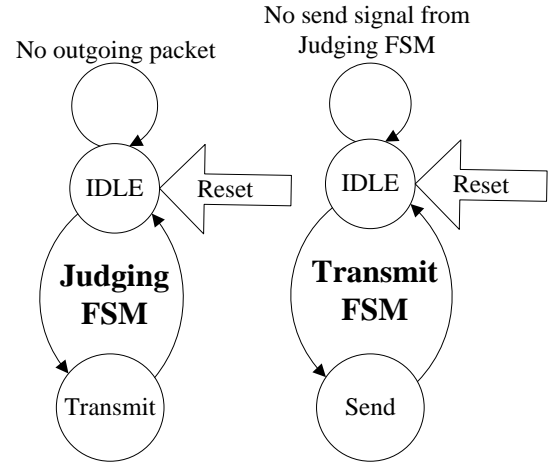

Figure 4. Simplified State Diagram of TX FSM.

### D. Timing Optimization

Design is following the principles listed as follows to achieve better timing performance.

*1) Synchronized Design:* Design is implemented in a synchronized way to use the large number of synchronous resources in FPGA.

*2) Avoiding Latches:* Design is implemented carefully to avoid any latches unwanted to appear. Statements such as "if-else" and "case" are examined to assure there are no uncompleted conditions. Initial value is assigned to every register.

*3) One-hot Code:* One-hot code is used in FSMs to achieve better timing performance.

*4) Pipelined Design:* Modules are designed to process data only when the requirements are met, not until the whole packet is obtained.

*5) Constraints:* Proper area and timing constraints are used to guide the tools.

### III. IMPLEMENTATION & VERIFICATION

All the modules introduced above as well as a Test module was coded in Verilog HDL, synthesized and implemented using ISE (IDE provided by Xilinx), and finally downloaded to a Xilinx ML507 Development Board. A standard PC system with PCI Express bus as well as Fedora 12 is used. Driver in Linux is developed to support applications to send or receive data.

Test module is attached to RX module and TX module. When a Write TLPs is received from RX module, the lower 8 bits of the first DW data will be used to driven 8 LEDs, When a Read TLPs is received from the RX module, Test module will send a Cpl TLP to TX module the lower 8 bits data of which is the value of switches on ML507.

### A. Resources Usage

The resources used by RX module, TX module are shown in Fig. 5. This result is estimated after synthesis.

Resources consumed by 4 FIFOs in TX module are included here.

*B. Verification*

All the modules introduced above as well as a Test module was coded in VerilogHDL, synthesized and implemented using ISE (IDE provided by Xilinx), and finally downloaded to a Xilinx ML507 Development Board.

The board which LEDs are driven by data 0xFF and 0xA5 given by CPU is shown in Fig. 6.

When applications of Fedora write and read the ML507 uninterruptedly, the TLPs received sent out by PWrapper are captured in Fig. 7 using Chipscope.

If another ML507 is used to write and read uninterruptedly, the TLPs are captured in Fig. 8.

The average data speed can achieve more than 1.8 Gbps (Giga bits per second) which is shown in Fig.8. The speed presented in Fig. 7 is far below this speed, because operating system rather than the design is the bottleneck here.

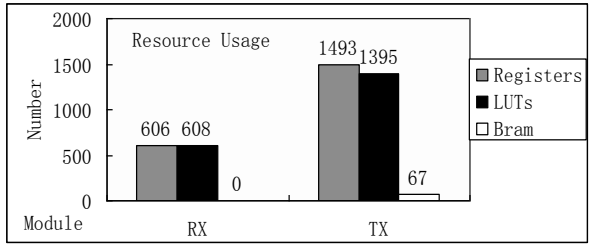

Figure 5.　Resources Usage.

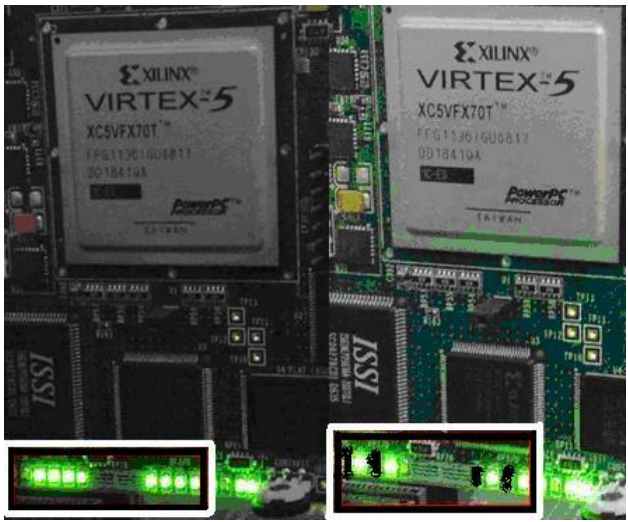

Figure 6.　LEDs on the developing board controlled by CPU.

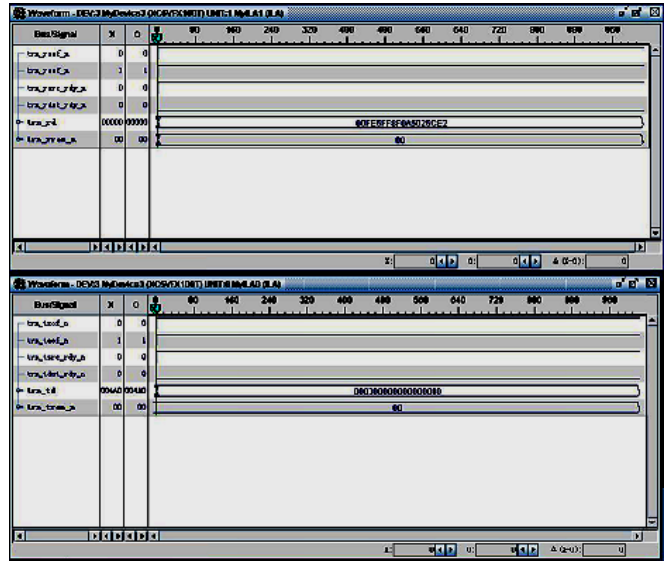

Figure 7.　Chipscope waveforms of write and read operations by CPU.

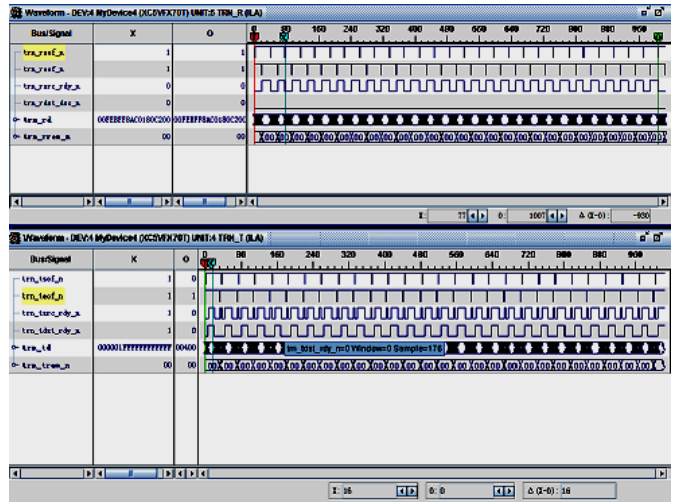

Figure 8.　Chipscope waveforms of write and read operations by another board.

IV. CONCLUSIONS

PCI Express becomes the standard interconnect in today's CPU centered systems for its own advantages. FPGA can provide fully integrated PCI Express solution. This paper proposes a IP core named PWrapper with simple but flexible FIFO interface, which provides an easier and faster way for engineers to complete a PCI-E based design.

PWrapper was implemented after being simulated on a FPGA board. Its resource demand was evaluated and results show that the performance is about 1.8Gbps. For higher rate of data transmission, DMA is suggested.

ACKNOWLEDGEMENT

The authors acknowledge the support of Important National Science & Technology Specific Projects (2011ZX03001-004-02) and National 863 project (60902049, 60873190).